\documentclass[fleqn,12pt,twoside]{article}
\usepackage{espcrc1}

\usepackage{graphicx}
\usepackage[figuresright]{rotating}

\newcommand{\AmS}{{\protect\the\textfont2
  A\kern-.1667em\lower.5ex\hbox{M}\kern-.125emS}}

\def\Qz{{\bf Q}}
\def\Q1{{\bf Q}_1}

\def\qj{{\bf q}_j}

\def\utm{{\bar u}_T}
\def\utms{\overline{u_T^2}}

\def\d{{\rm d}}

\def\e{{\rm e}}

\def\oV{\overline V}

\def\ee{e$^+$e$^-$}
\def\ss{${\rm s}\bar{\rm s}\;$}
\def\ssb{\langle {\rm s}\bar{\rm s}\rangle}
\def\uub{\langle {\rm u}\bar{\rm u}\rangle}
\def\ddb{\langle {\rm d}\bar{\rm d}\rangle}

\title{Statistical hadronisation phenomenology}

\author{F. Becattini\address{Universit\`a di Firenze and INFN Sezione di Firenze, \\ 
        Via G. Sansone 1, I-50019, Sesto F.no, Firenze, Italy.}}
       
\begin{document}

% typeset front matter
\maketitle

\begin{abstract}

The analyses of hadron production in the framework of the statistical 
hadronisation model are reviewed. The analysis of average multiplicities in 
collisions at relatively low centre-of-mass energy confirms previous findings, 
namely the universality of hadronisation temperature and of strange to 
non-strange quark production rate. The study of transverse momentum spectra of 
identified hadrons allows a further determination of the hadronisation 
temperature which is found to be compatible with that obtained from fits 
to average multiplicities.   

\end{abstract}

%***********************************************************************
\section{INTRODUCTION}
%***********************************************************************

The idea of a statistical approach to hadron production in high energy 
collisions dates back to '50s \cite{fermi} and '60s \cite{hage} and 
it has been recently revived by the observation that hadron multiplicities 
in \ee and pp collisions agree very well with a thermodynamical-like ansatz 
\cite{beca1,beca2,beca3}. This finding has also been confirmed in hadronic 
collisions and it has been interpreted in terms of a pure statistical filling 
of multi-hadronic phase space of assumed pre-hadronic clusters (or fireballs) 
formed in high-energy collisions, at a critical value of energy density 
\cite{beca3,heinz,stock}. In this framework, temperature and other thermodynamical 
quantities have a purely statistical meaning and do not involve the existence 
of a hadronic thermalisation process through multiple collisions on an 
event-by-event basis. 

So far, this proposed statistical cluster hadronisation model has been 
mainly tested against measured abundances of different hadron species 
for a twofold reason. Firstly, unlike momentum spectra, they are quantities 
which are not affected by hard (perturbative) QCD dynamical effects but are 
only determined by the hadronisation process; indeed, in the framework of
a multi-cluster model, they are Lorentz invariant quantities which are 
independent of individual cluster's momentum. Secondly, they are fairly easy 
to calculate and provide a very sensitive test of the model yielding an 
accurate determination of the temperature. It is now a quite natural step
to test further observables and to assess their consistency with the results 
obtained for multiplicities. One of the best suited observables in this regard 
is the transverse momentum of identified hadrons (where transverse is meant 
to be with respect to beam line in high energy hadronic collision, and 
thrust or event axis in high energy \ee collisions) because, amongst all 
projections of particle momentum, this is the most sensitive to hadronisation 
or, conversely, the least sensitive to perturbative QCD dynamics.

In this paper, after a brief sketch of the statistical hadronisation model,
the results obtained in a comprehensive analysis of average 
multiplicities and transverse momentum spectra of identified hadrons in 
various elementary collisions are summarized. A more detailed description 
of both the model and the analysis can be found in ref.~\cite{becapt}.

%*************************************************************************
\section{HADRON MULTIPLICITIES}
%*************************************************************************

In order to obtain quantitative predictions in the model, the assumption 
of local statistical equilibrium is not enough because average
values of physical observables also depend on clusters configurations
in terms of volume, four-momentum and charges. Thus, the probabilities of
clusters configurations are needed, but they are governed by the particular
dynamical evolution of the process which is beyond the statistical
ansatz. For Lorentz-invariant observables such as average multiplicities,
one can take advantage of their independence of cluster momenta and try to
establish a mathematical equivalence between the actual system of clusters
and one {\em equivalent global cluster} (EGC) with volume and quantum numbers
equal to the sum of single volumes and quantum numbers. This equivalence
would lead to a dramatic reduction of the number of model's free parameters and 
one would be essentially left with volume and mass of the EGC. However, for this
equivalence to apply, a special form of cluster masses and charges distribution
for fixed volumes is required \cite{becapt}. If this is the case, the average
multiplicity of the $j^{\rm th}$ hadron reads:

\begin{equation}\label{mult}
\langle n_j \rangle = \int \d M \int \d V \; \chi(M,V) \; n_j(M,\Qz,V) 
\end{equation}
where $n_j(M,\Qz,V)$ is the average multiplicity for fixed EGC's mass, quantum
numbers $\Qz=(Q_1,\ldots,Q_n)$ and volume and $\chi$ is the EGC's mass and volume
distribution. If the EGC is sufficiently large, the canonical ensemble can be
introduced through a saddle-point approximation and the above multiplicity 
can be written as: 

\begin{equation}\label{mult2}
\langle n_j \rangle = \int \d T \int \d V \; \zeta(T,V) \; n_j(T,\Qz,V) \simeq 
n_j(\overline T,\Qz,\overline V)  
\end{equation}  
where $\overline T$ and $\overline V$ are meant to be mean values independent
of hadron species. The last equality is a good approximation if $\zeta$ is a 
peaked function of $T$ and $V$. It must be stressed that, in this procedure, 
temperature can be a well defined concept only in a global sense, i.e. for the 
EGC, while individual clusters might be so small that a microcanonical 
description is essential. Otherwise stated, $T$ could be introduced at a global 
level even though not locally defined. 

In order to reproduce the experimentally measured multiplicities, an extra 
strangeness suppression parameter must be introduced. This has been usually 
done \cite{beca1,beca2,beca3} by means of a factor $\gamma_S^{n_S}$ (where $n_S$ is 
the number of strange quarks in the hadron) multiplying, in the Boltzmann limit, 
the full-equilibrium primary multiplicity on the right-hand-side of eq.~(\ref{mult2}). 
For some of the considered colliding systems, a new parametrisation has been 
used in which the number of newly produced ${\rm s} + \bar{\rm s}$ constituent 
quarks is considered as an additional charge to be fixed into the final 
hadrons. The number of produced \ss pairs are allowed to fluctuate poissonianly 
around a mean value $\ssb$, taken as a further free parameter, so that the primary 
average multiplicity actually reads:
 
\begin{equation}\label{mult3}
 \langle n_j\rangle^{\rm primary} = \frac{V T(2J_j+1)}{2\pi^2} \sum_{K=0}^\infty
 \frac{\e^{-\ssb}\ssb^K}{K!} \sum_{n=1}^\infty (\mp 1)^{n+1}\;\frac{m^2}{n}\;
 {\rm K}_2\left(\frac{n m}{T}\right)\, \frac{Z(\Qz-\qj)}{Z(\Qz)}
\end{equation} 
where $\Qz = (Q,N,S,N_S)$ and $\qj = (Q_j,N_j,S_j,N_{Sj})$ are the quantum 
numbers vectors with components electric charge, baryon number, strangeness and
number of strange quarks associated with the initial state and with the $j^{\rm th}$ 
hadron respectively and the $Z$ are the canonical partition functions. The 
free parameters in Eq.~(\ref{mult3}) are the temperature $T$,
the volume $V$ and the mean number of newly produced $\ssb$. From these parameters,
one can calculate the Wroblewski parameter $\lambda_S = 2\ssb /(\uub + \ddb)$,
the ratio between newly produced \ss pairs and light quark pairs by using the
fitted primary multiplicities of all the hadron species.   

\begin{figure}[htb]
\begin{minipage}[t]{77mm}
\includegraphics[width=18pc]{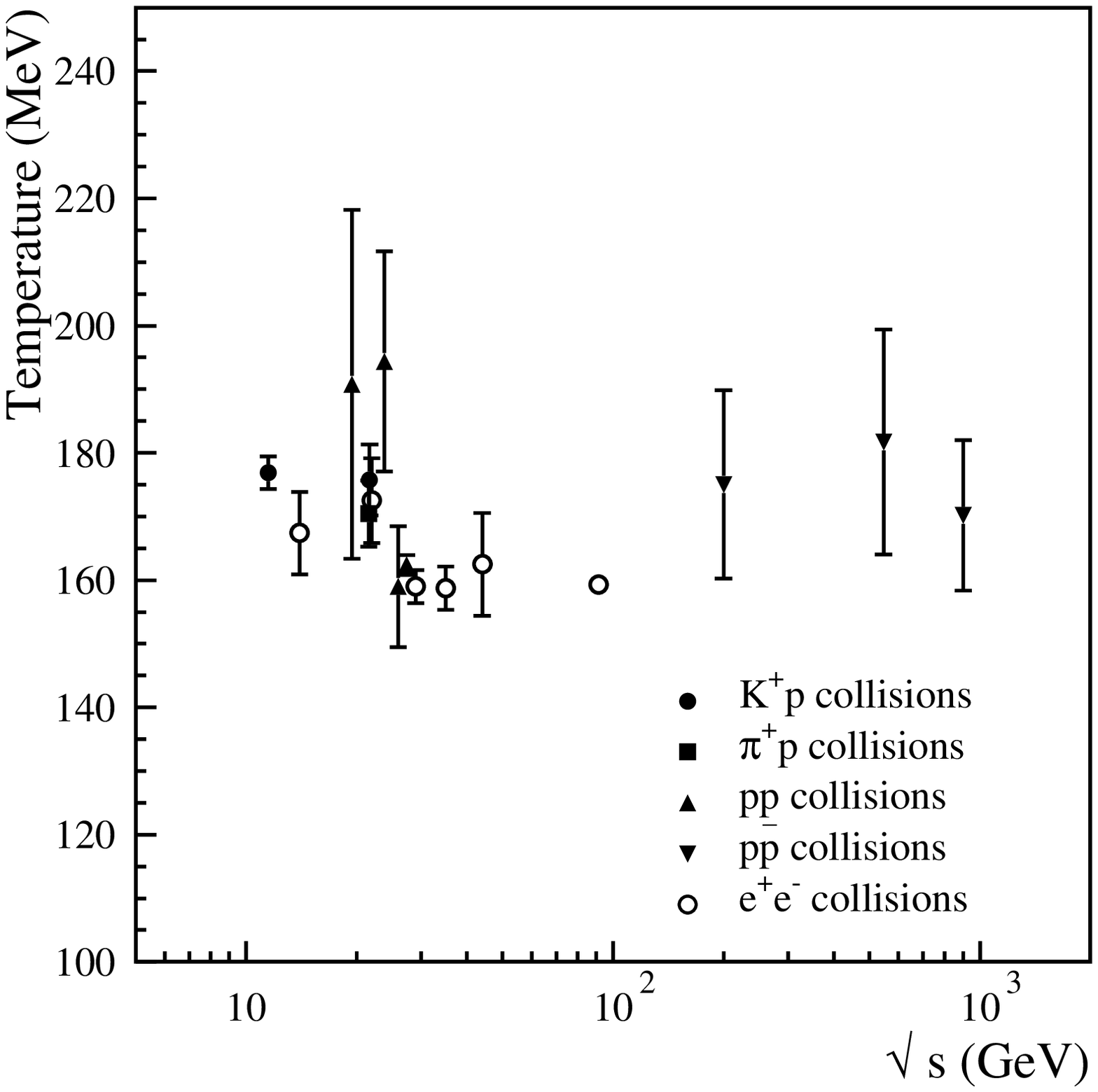}
\caption{Temperatures fitted with hadron multiplicities ~\cite{beca3,becapt} 
in high energy collisions. The \ee point at $\sqrt s = 91.2$ GeV has been 
calculated for this conference.}
\label{temp}
\end{minipage}
%
%\hspace{\fill}
%
\begin{minipage}[t]{77mm}
\includegraphics[width=18pc]{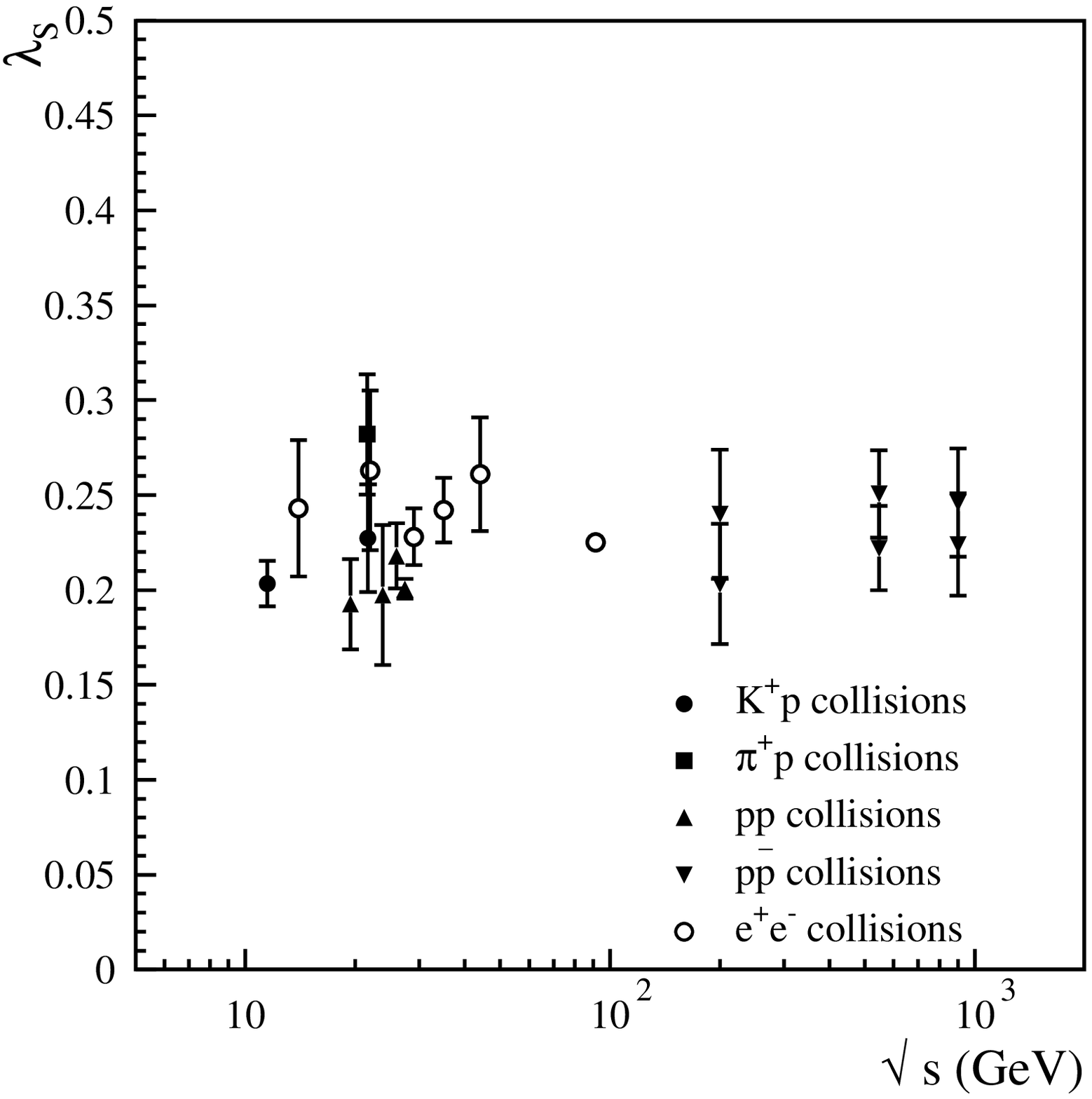}
\caption{Wroblewski factor $\lambda_S$ calculated with fitted primary hadron
multiplicities~\cite{beca3,becapt}. The \ee point at $\sqrt s = 91.2$ GeV has 
been calculated for this conference.}
\label{ls}
\end{minipage}
\end{figure}

The results of the fit to average multiplicities are summarized in figs.~\ref{temp}
and \ref{ls} showing the best-fit values of $T$ and $\lambda_S$ for several 
kinds of collisions. The striking feature of both plots is the fair constancy of the 
quantity over two orders of magnitude of centre-of-mass energy regardless of the 
initial colliding particles. Indeed, $T$ seems to rise as centre-of-mass energy 
decreases, but the genuiness of this effect is not established yet.    
%-------------------------------------------------------------------------- 
%---------------------------------------------------------------------------

%**********************************************************************
\section{STUDY OF TRANSVERSE MOMENTUM SPECTRA}
%**********************************************************************

The main goal of the analysis of transverse momentum spectra of identified hadrons
is the assessment of their consistency with the prediction of the statistical hadronisation
model. This has a twofold implication: firstly, spectrum shapes of different hadrons 
at a given centre-of-mass energy should be described by essentially the same 
parameters; secondly, the best-fit $T$'s should be in agreement with the temperature 
fitted with hadron multiplicities. In view of this objective, a natural requirement 
for the data set for a given collision and centre-of-mass energy, is the existence of 
a considerably large sample of both measured transverse momentum spectra and integrated 
multiplicities of different hadron species. Furthermore, centre-of-mass energy must be 
high enough to allow the use of a canonical formalism, what is expected to occur above 
roughly $\sqrt s \approx 10$ GeV \cite{becapt}. Essentially, four collision systems 
fulfilling these requirements have been found: 
K$^+$p at $\sqrt s = 11.5$ and $\sqrt s = 21.7$ GeV, $\pi^+$p at $\sqrt s = 21.7$ 
GeV and pp at $\sqrt s = 27.4$ GeV.

Provided that the EGC exists and that the average transverse (with respect to beam 
line in hadronic collisions and event axis in \ee collisions) four-velocity $\utm$ 
of clusters is sufficiently small, the transverse momentum spectrum of the 
$j^{\rm th}$ hadron species can be written, in the canonical ensemble, as:

\begin{eqnarray}\label{ptspec} 
&& \Big\langle \frac{\d n_{j}}{\d p_T} \Big\rangle = \frac{\oV (2J_j+1)}
 {2\pi^2 \sqrt{1+\utms}} \sum_{n=1}^\infty (\mp 1)^{n+1} \, m_T \, p_T 
 {\rm K}_1 \left( \frac{n \sqrt{1+\utms}\,m_T}{T}\right) 
 {\rm I}_0 \left( \frac{n \utm p_T}{T}\right) \nonumber \\
&& \times \frac{Z(T,\Qz-n\qj,\oV)}{Z(T,\Qz,\oV)}
 + \sum_k \Big\langle \frac{\d n_{j}}{\d p_T} \Big\rangle^{k \rightarrow j}
\end{eqnarray} 
The first term on the right hand side is the primary spectrum while the second term
is the contribution to the spectrum due to the decays of hadron $k$ into the hadron
$j$ either directly or through intermediate steps. The latter has been calculated 
by means of a newly proposed mixed analytical-numerical method \cite{becapt}. 
%----------------------------------------------------------------------
\begin{figure}[htb]
\begin{minipage}[t]{77mm}
\includegraphics[width=18pc]{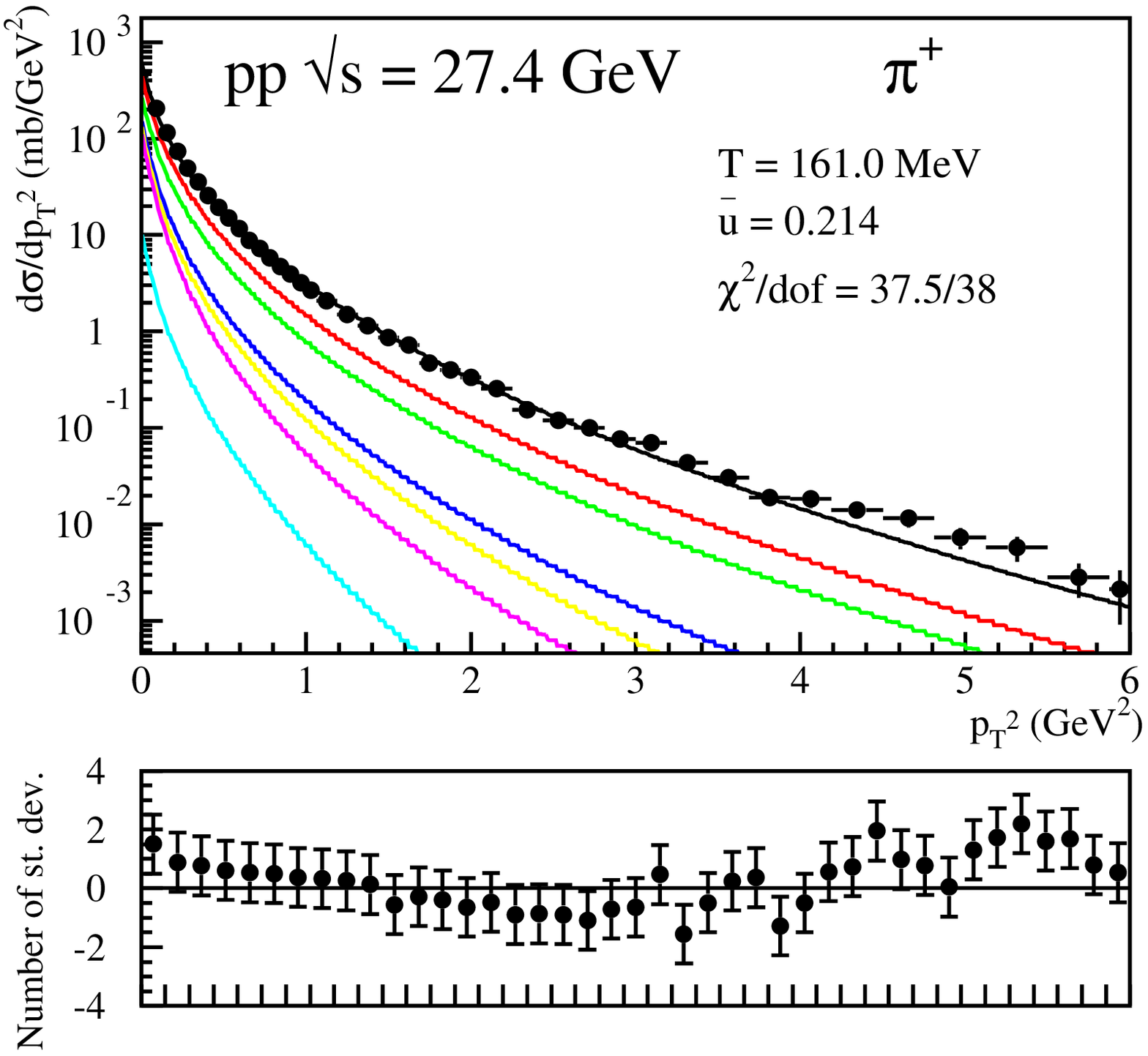}
\caption{Fitted $\pi^+$ transverse momentum spectrum in pp collisions at $\sqrt s= 27.4$
GeV. The grey-scaled solid lines are the cumulative contributions of the decays of
various sets of parent hadrons. Below: residuals distribution.}
\label{spectrum}
\end{minipage}
%
%\hspace{\fill}
%
\begin{minipage}[t]{77mm}
\includegraphics[width=18pc]{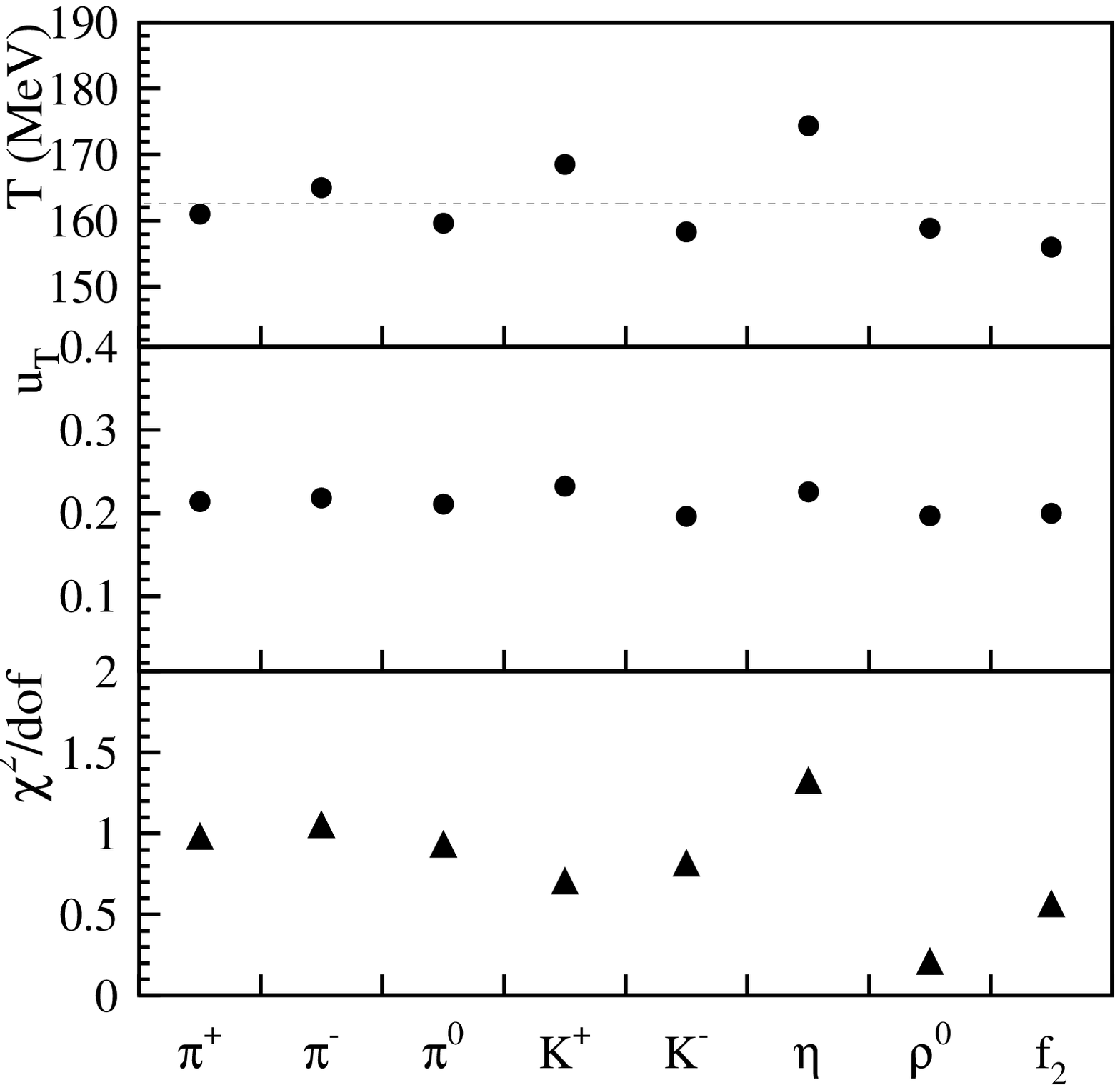}
\caption{Fitted temperatures and cluster average transverse four-velocities in
pp collisions at $\sqrt s= 27.4$ GeV with identified hadrons spectra. Also 
shown the minimum $\chi^2$ values.}
\label{summapp}
\end{minipage}
\end{figure}
%-----------------------------------------------------------------------

The analysis is described in full detail in ref.~\cite{becapt}; hereby, the main 
results are briefly summarized. The free parameters in the formula~(\ref{ptspec}) 
are essentially $T$ and $\utm$, the others being fixed by the multiplicity fit. 
For each identified hadron, there are several $\chi^2$ local minima located along a 
band in the $T-\utm$ plane. Thus, in order to single out a best-fit solution, the
minimum nearest to the {\em chemical} temperature (i.e. fitted with multiplicities)
$1\sigma$ band is chosen. The use of this method does not allow to give a conclusive 
answer to the question whether there is independent consistency between the 
temperatures governing the $p_T$-slopes and multiplicities respectively. Still,
it is possible to test the universality of $T$ and $\utm$ for different hadron 
species at a given centre-of-mass energy. This is indeed shown in fig.~\ref{summapp}
demonstrating that the fitted parameters for eight different particles in pp 
collisions are found to be in good agreement with each other. On the other hand,
some difficulties have been found in a similar analysis of hadrons measured in
K$^+$p and $\pi^+$ collisions at $\sqrt s = 21.7$ GeV.      

%**********************************************************************
\section{CONCLUSIONS}
%**********************************************************************

A comprehensive analyisis of multiplicities and transverse momentum spectra 
of identified hadrons in several high energy collisions confirms the agreement
of the data with the predictions of statistical hadronisation model. 
Particularly, the temperature estimated with these two observables 
are fully compatible with each other, which is an indication in favour of 
one of the key predictions of the model. The extracted temperature 
is fairly constant throughout with a value of about $\simeq 160$ MeV in 
the high energy limit, which is amazingly close to that determined in similar 
analyses in heavy ion collisions \cite{becahi}. The universality of $T_{\rm had}$ 
and its very value, close to the expected QCD critical temperature for 
deconfinement and chiral symmetry restoration, strongly suggest that 
hadronisation itself is a critical phenomenon occurring at a peculiar value 
of (local) energy density \cite{beca3,heinz,stock}. The production ratio of s 
quark with respect to light quark is fairly constant as well, with a value 
around $\approx 0.23$. 

%**********************************************************************

%**********************************************************************

\end{document}